\documentclass[conference,10pt]{IEEEtran}
\IEEEoverridecommandlockouts
\usepackage[utf8]{inputenc}
\usepackage{cite}
\usepackage{amsmath,amssymb,amsfonts}
\usepackage{algorithmic}
\usepackage{graphicx}
\usepackage{textcomp}
\usepackage[table]{xcolor}
\usepackage{tabularx}
\usepackage{subcaption}
\usepackage[font=scriptsize]{caption}

\newcolumntype{C}{>{\centering\arraybackslash}X} 

\def\BibTeX{{\rm B\kern-.05em{\sc i\kern-.025em b}\kern-.08em
    T\kern-.1667em\lower.7ex\hbox{E}\kern-.125emX}}
\begin{document}

\title{When GenAI Meets Fake News: Understanding Image Cascade Dynamics on Reddit}

\author{\IEEEauthorblockN{Saumya Chauhan}
\IEEEauthorblockA{
\textit{Computing + Mathematical Sciences} \\
\textit{California Institute of Technology}\\
Pasadena, California, USA \\
saumya.s.chau@gmail.com}
\and
\IEEEauthorblockN{Mila Hong}
\IEEEauthorblockA{
\textit{Computing + Mathematical Sciences} \\
\textit{California Institute of Technology}\\
Pasadena, California, USA \\
milazoehwork@gmail.com}
\and
\IEEEauthorblockN{Maria Vazhaeparambil}
\IEEEauthorblockA{\textit{Computing + Mathematical Sciences} \\
\textit{California Institute of Technology}\\
Pasadena, California, USA \\
maria.j.vazhaeparambil@gmail.com}
}

\maketitle

\begin{abstract}
AI-generated content and misinformation are increasingly prevalent on social networks. While prior research primarily examined textual misinformation, fewer studies have focused on visual content's role in virality. In this work, we present the first large-scale analysis of how misinformation and AI-generated images propagate through repost cascades across five ideologically diverse Reddit communities. By integrating textual sentiment, visual attributes, and diffusion metrics (e.g., time-to-first repost, community reach), our framework accurately predicts both immediate post-level virality (AUC=0.83) and long-term cascade-level spread (AUC=0.998). These findings offer essential insights for moderating synthetic and misleading visual content online.
\end{abstract} 

\section{Introduction} 

The rise of social media platforms like Reddit, Instagram, and X has transformed information dissemination. Unlike traditional media, which rely on editorial oversight to ensure accuracy, these platforms prioritize user-driven engagement signals, such as upvotes and shares, over credibility. Consequently, physiologically arousing or false content often spreads faster than neutral or truthful information \cite{berger2012makes, vosoughi2018spread}.

Compounding this issue is the rise of generative AI (GenAI) and deepfake image manipulation tools. Visual content is processed more quickly than text and often transcends language barriers. Past research has also found that images are inherently more memorable than text \cite{standing1973learning, isola2011makes}. Regardless, distinguishing real from synthetic visuals remains difficult, even with recent advances in detection \cite{farid2009image,cozzolinoClipDetection2023}. While recent studies examine visual misinformation, such as COVID-19 images on Twitter \cite{wang2023imagesCOVID} and meme virality on Reddit \cite{sah2025decoding}, research on virality in ideologically driven communities remains sparse. 

To address this, we analyze how misinformation and GenAI influence the virality of visual and textual posts across diverse Reddit communities. We specifically study virality at two distinct levels: immediate individual post popularity ("post-level virality") and long-term content propagation through sharing or reposting ("cascade-level virality"). A diffusion cascade refers to the complete pathway through which a post spreads across a network via repeated reposting. At the cascade level, we examine features of entire diffusion cascades, such as overall size, structural complexity, and community reach.

Specifically, we explore the following questions:

\begin{enumerate}
\item How do visual and textual features affect virality at the individual post and cascade levels?

\item When predicting how widely content spreads (cascade-level virality), how do content-based predictors such as text and visual attributes compare against early sharing signals and community interaction patterns (diffusion context-based features)?

\item What post-specific features characterize pure misinformation, pure AI-generated, and misinformation and AI-generated imagery ("mixed-flag" content) classifications
at the post and cascade level?
\end{enumerate}

The first question builds upon findings that emotionally engaging or misleading content tends to be more viral \cite{berger2012makes, vosoughi2018spread}. Prior studies have focused on general imagery, typically without distinguishing misinformation or synthetic media \cite{khosla2014makes, dezaImageVirality2015}. By explicitly examining misinformation and AI-generated visuals, we offer targeted insights into how these specific content types shape virality uniquely within ideologically driven communities.

The second question addresses the gap in understanding Reddit's diffusion dynamics due to its lack of an explicit social graph—unlike Twitter, where propagation is more straightforward to analyze. Our study reconstructs repost cascades by leveraging shared image URLs and crossposting data, allowing us to compare how intrinsic features of posts (content-based) versus early sharing behaviors and diffusion metrics (context-based) predict virality. Additionally, we introduce novel visual markers specific to generative AI, such as digital artifacts and image noise, identifying previously unexplored connections between these visual elements and content diffusion.

The third question investigates the distinct and combined effects of misinformation and AI-generated imagery ("mixed-flag posts"). Mixed-flag posts—those identified as containing both misinformation and synthetic or manipulated imagery—may drive unique engagement patterns that neither misinformation nor synthetic visuals alone can explain. Prior research highlights misinformation's power to engage via novelty or emotional provocation, and AI-generated content's effectiveness through visually striking imagery and positive framing despite questionable credibility \cite{vosoughi2018spread,drolsbachMisinformation2025}. However, how these features interact in combination remains under-explored. We address this gap by systematically studying pure misinformation, pure AI-generated content, and mixed-flag posts, providing nuanced insights into their distinct engagement patterns and diffusion trajectories.

\subsection{Our Contribution}

We present the first large-scale cascade-level study of visual content diffusion on Reddit, focusing on how misinformation and GenAI content spread across user communities. Our dataset includes 1,800+ repost cascades clustered via shared image URLs and crossposts, encompassing 5,660 posts from five ideologically diverse subreddits. Each cascade is enriched with multimodal features: text sentiment, visual descriptors (e.g., digital manipulation indicators, image sharpness, and GenAI-specific markers), and user interaction metrics (e.g., upvote ratio, comment counts). Diffusion-based attributes such as cascade depth, Wiener index (structural virality), time-to-first repost, and cross-subreddit reach are also computed. Misinformation posts receive unusually high engagement (mean score of 36,000+, 2,400+ comments), while GenAI images show broader subreddit spread (1.106 average cross-community reach). Mixed-flag cascades (misinfo + AI) exhibit the strongest dynamics, with an average size of 26.96, depth of 25.96, and structural virality of 7.14.

Our framework effectively supports both cascade-level and post-level virality predictions through early diffusion signals combined with multimodal post features. At the post-level, visual features dominate predictions, with markers like image noise and digital artifacts significantly influencing virality (post-level prediction accuracy 87\%, AUC = 0.83). At the cascade-level, our best model integrating multimodal content and early diffusion achieves an AUC of 0.998, surpassing both content-only (0.957) and diffusion-only (0.995) baselines. The robustness of our GenAI and misinformation pseudolabeling methods, validated through manual reviews and cross-dataset evaluations, significantly enhances our model’s predictive accuracy and reliability. By bridging content-based analysis with diffusion modeling, we provide a comprehensive empirical framework for understanding multimodal misinformation and synthetic media propagation on Reddit.

\section{Methods}
To investigate how visual misinformation and GenAI content spread across Reddit, we developed a multimodal analysis pipeline targeting five visually driven subreddits, r/conspiracy, r/politics, r/conservative, r/propagandaposters, and r/deepfakes. Our approach includes dataset filtering, misinformation and GenAI classification, repost cascade modeling, and predictive analysis. This enables us to examine early indicators of virality, diffusion dynamics of harmful content, and multimodal predictors of spread.

\begin{figure}[htbp]
\centerline{\includegraphics[width=1.0\linewidth]{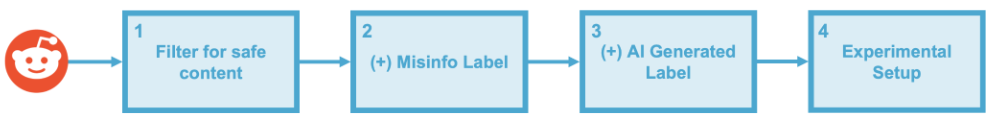}}
\caption{Overall Architecture Pipeline}
\label{fig:pipeline}
\end{figure}

\subsection{Data Collection \& Labeling}
\subsubsection{Safe Search}
To ensure the appropriateness of our dataset, we implemented an automated filtering process to exclude any images classified as not safe for work (NSFW). We used a pretrained image classification model from the Hugging Face Transformers library prior to inclusion in our analysis due to its accuracy and scalability after exploring several alternatives, including other pretrained models, NSFW Python libraries, and a custom Tensor Flow model. Manual review of 100 images (50 safe, 50 NSFW) showed ~95\% alignment with human judgment, confirming the model’s reliability.

\subsubsection{Misinformation}

To assess textual misinformation, we fine-tuned a classifier on the Fakeddit dataset consisting of over 110,000 images from Reddit. We selected this method due to the relevance of this data and its improved performance over several baseline models, including zero-shot classifiers based on large language models (e.g., BART, RoBERTa) and sentiment-based heuristics. To validate the accuracy of our fine-tuned model, we conducted a manual review of 100 random post titles, and we observed about 79.5\% agreement between model outputs and human judgments.

\subsubsection{GenAI}
To develop an AI-generated pseudolabel for each image, we used the pre-trained ClipBased-SyntheticImageDetection model introduced by \cite{cozzolinoClipDetection2023} to extract image embeddings followed by a SVM classifier with a RBF-kernel. We chose this model because of its ability to generalize across different generative models with only few-shot learning \cite{cozzolinoClipDetection2023}. To train, we curated a labeled dataset by combining real and AI-generated images from diverse sources. A key challenge was the lack of large-scale, ground-truth-labeled datasets for social media content, which often includes noisy formats like memes, screenshots, and annotated images. We experimented with multiple dataset combinations to assess their effect on generalization, especially on out-of-sample AMMeBa validation. Our final dataset included:

\begin{table}[ht!]
\centering
\caption{Training data details of real vs. fake content.}
\begin{tabular}{|l|r|r|}
\hline
\rowcolor[HTML]{E6E6FA}
\textbf{Training Data} & \textbf{Real} & \textbf{Fake} \\
\hline
AI Generated Images vs. Real Images Dataset \cite{zhangAIDataset} & 10,000 & 2,000 \\
Google Research’s AMMeBa Dataset \cite{dufour2024AMMeBa} & 219 & 219 \\
Manually Labeled Reddit Data & 202 & 202 \\
\hline
\textbf{Total} & \textbf{10,421} & \textbf{2,421} \\
\hline
\end{tabular}
\label{tab:training_data}
\end{table}

By combining scale \cite{zhangAIDataset}, domain relevance \cite{dufour2024AMMeBa}, and platform-specific samples (Reddit), we constructed a dataset optimized for performance on both typical and noisy social media images. We selected a lightweight classifier by evaluating six models—Random Forest (RF), Logistic Regression, MLP, SVM (linear \& RBF), and XGBoost—on 109 out-of-sample AMMeBa images. We chose an RBF-kernel SVM for its superior accuracy (91\%) and F1-score.

\subsection{Feature Extraction \& Virality Metric}

We extracted multimodal features, textual, visual, and metadata, for each Reddit post, computed a Virality Attention Index (VAI), and constructed content cascades to measure structural virality.

\subsubsection{Textual Features (Post-Level)}

We extract text-based features to capture linguistic style, sentiment, and clickbait tendencies. \emph{Title and caption embeddings} are generated via a pretrained BERT model, producing a 768-dimensional vector that encodes semantic and contextual information. \emph{Sentiment scores} are computed using VADER, yielding compound (overall polarity in \([-1,+1]\)), positive, and negative values. To assess \emph{clickbait cues}, we include the length and flag presence of keywords like “breaking,” “viral,” and “shocking,” which often signal sensational or high-engagement content.

\subsubsection{Visual Features (Post-Level)}

To characterize each post's image thumbnail, we extract visual features capturing quality, content, and dimensions. We compute noise, variance, and error scores to assess sharpness, compression, and potential manipulation. Using YOLOv5, we detect objects (e.g. ``people'', ``animal,'' ``car,'' etc.) as binary flags. We log thumbnail width and height to capture aspect ratio and resolution. Finally, we include two image-level flags from Section 3.4’s classifiers: \texttt{misinfo\_flag (image)} (likely manipulated) and \texttt{genai\_flag (image)} (likely AI-generated), which help identify fabricated visuals and estimate attention potential.

\subsubsection{Metadata \& User-Interaction Features}

We extract metadata and interaction features reflecting post performance and author behavior. These include \emph{upvote ratio} (upvotes / (upvotes + downvotes)), \emph{total awards}, and \emph{explicit crossposts} (\texttt{num\_crossposts}). We record whether the post is original content (Reddit flag) and the \emph{hour of posting} (\texttt{hour}) to capture temporal context. A binary \emph{is\_crossposted} flag marks posts shared across subreddits. Finally, intensity relative to voting is quantified using the \emph{engagement ratio} (\(\frac{\text{Total\_comments}}{ \max(\text{Score},1)}\)).

\subsubsection{Virality Attention Index (VAI)}

To measure early-stage virality, we define the Virality Attention Index (VAI), a scalar metric balancing engagement with post recency:
\[
  \text{VAI} 
  \;=\;
  \frac{\text{Score} \;+\; \alpha \times \text{Total\_comments}}
       {\bigl(\text{age\_hours} + \tau \bigr)^{\beta}}.
\]
\begin{itemize}
  \item \(\text{Score}\) is the number of upvotes
  \item \(\text{Total\_comments}\) is the number of comments
  \item \(\text{age\_hours} = \) (current timestamp - post timestamp) in hrs
  \item \(\alpha = 1,\,\beta = 1,\,\tau = 1\) (chosen to avoid division by zero)
\end{itemize}

VAI captures the “early burst” effect: high values reflect rapid early engagement, while lower values indicate slower or stale posts. It is used both as a virality label (top 20\% vs.\ others) and as a feature in downstream models for detecting misinformation and AI-generated content.

\subsubsection{Cascade Construction (DSU Clustering)}

To group related posts sharing content, we apply a Disjoint‐Set Union (DSU) clustering procedure based on:
\begin{enumerate}
\item \textbf{Image URL Matching:} Posts with identical image URLs are merged.
\item \textbf{Crosspost Links:} Posts with a \texttt{crosspost\_parent} are unioned with the parent’s set.
\item \textbf{Same Author Reposts:} Posts with visually similar content by the same author are merged.
\end{enumerate}
This yields disjoint \emph{cascade components} \(\{C_1, C_2, \ldots\}\), each representing a group of related posts across subreddits. We discover 1,804 such cascades and compute per-cascade statistics: total upvotes, number of subreddits, average repost delay, and text/image content entropy.

\subsubsection{Canonical Repost Graph (For Each Cascade)}

For each cascade \(C\), we construct a time-respecting repost graph \(G_C\):
\begin{enumerate}
\item \textbf{Order Posts:} Sort posts by timestamp.
\item \textbf{Assign Parent Edges:} If an explicit crosspost parent \(p_i\) exists in 
\(C\), link parent \(p_j\)
to \(p_i\). Else, link \(p_i\) to its immediate predecessor in time.
\end{enumerate}
The resulting directed graph \(G_C\) encodes both repost lineage and temporal flow.

\subsubsection{Structural Virality (Wiener Index)}

To measure cascade shape, we find the structural virality of each repost graph \(G_C\):
\begin{enumerate}
\item \textbf{Undirected Form:} Convert \(G_C\) to an undirected graph 
  \(
    U_C \;=\; G_C.\texttt{to\_undirected()}.
  \)
\item \textbf{Shortest Paths:} For every pair of distinct nodes \((u, v)\) in \(U_C\), compute the shortest‐path length \(d(u, v)\).
\item \textbf{Wiener Index:}
To distinguish between broadcast‐style cascades (many direct reposts of a single root) and chain‐style cascades (deep, multi‐generation propagation), we compute the structural virality (also known as the Wiener index) for each repost graph \(G_C\):
\[
  \mathrm{StructuralVirality}(C)
  \;=\;
  \frac{1}{\binom{|C|}{2}}
  \sum_{u < v} d(u, v)\,.
\]
\end{enumerate}
Low virality ($\approx1$) indicates a star-like broadcast; high values reflect deeper, chain-like propagation. This metric captures both cascade size and how content diffused—via broad hubs or sequential reposting. 
Using post-level features and VAI, these cascade-level signals enable robust modeling of early virality, misinformation, and generative content detection.

\subsubsection{Feature Importance via SHAP Values}
To interpret model predictions and quantify the influence of individual features, we compute SHAP (SHapley Additive exPlanations) values, a unified measure of feature importance. SHAP values offer consistent and locally accurate attribution scores, helping identify which features most strongly drive virality predictions. We use these insights in Section 3 (Experimental Results) to interpret feature contributions clearly and intuitively.

\section{Experimental Results}
\subsection{Experimental Setup}
\begin{figure}[htbp]
\centerline{\includegraphics[width=1.0\linewidth]{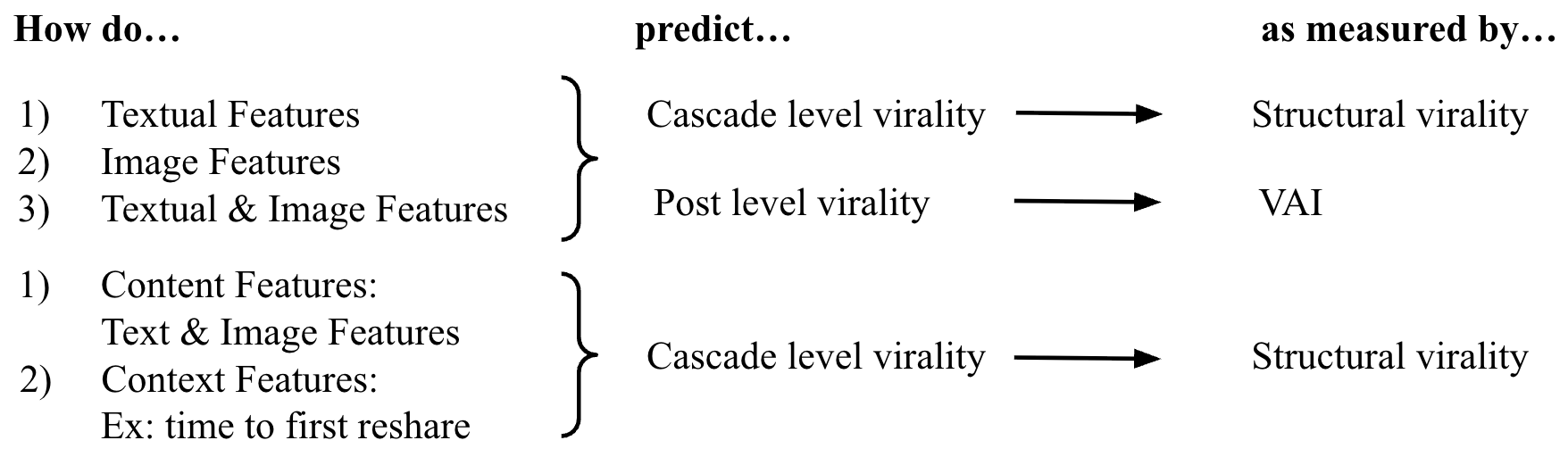}}
\caption{Experimental Setup Overview}
\label{fig:setup}
\end{figure}

We conduct eight experiments to evaluate how different feature sets predict cascade and post level virality. Across all experiments, we evaluate four models: LightGBM, RF, RF with Extra Trees, and Gradient Boosting.

\begin{figure*}[t]
  \centering
  \begin{subfigure}[b]{0.32\textwidth}
    \includegraphics[width=\textwidth]{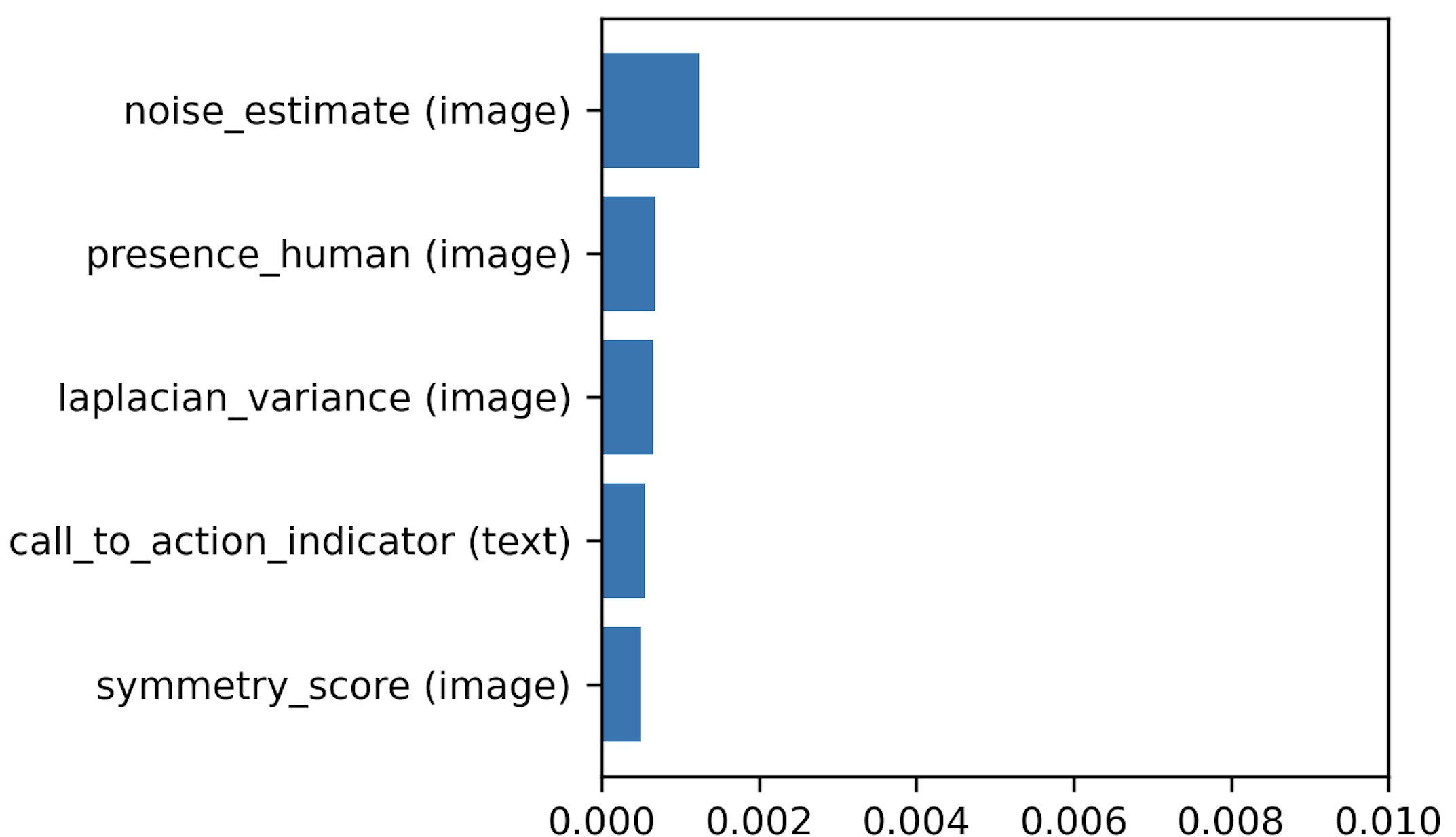}
    \caption{}
    \label{fig:rq1_post}
  \end{subfigure}
  \hfill
  \begin{subfigure}[b]{0.32\textwidth}
    \includegraphics[width=\textwidth]{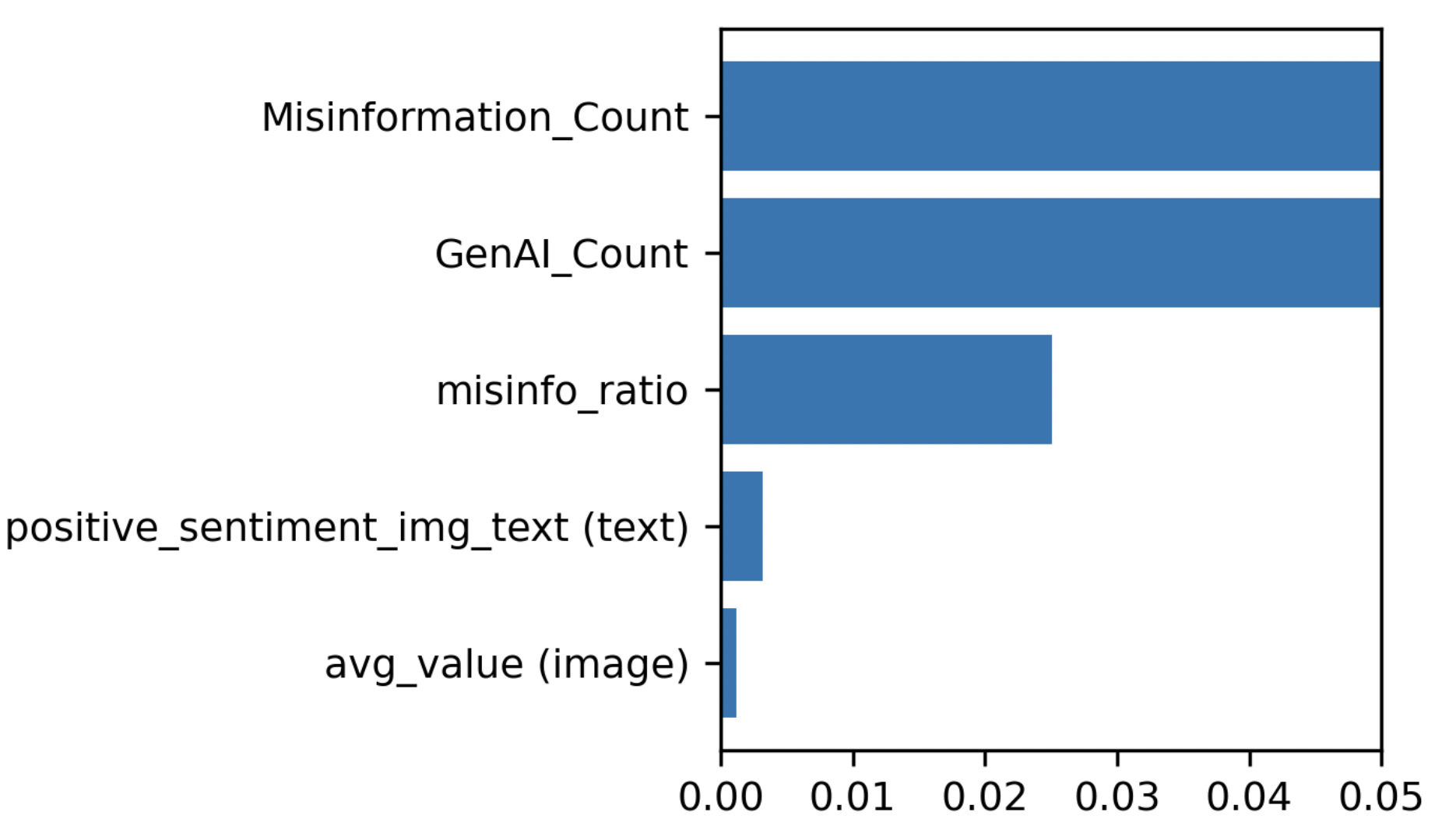}
    \caption{}
    \label{fig:rq1_cascade}
  \end{subfigure}
  \hfill
  \begin{subfigure}[b]{0.32\textwidth}
    \includegraphics[width=\textwidth]{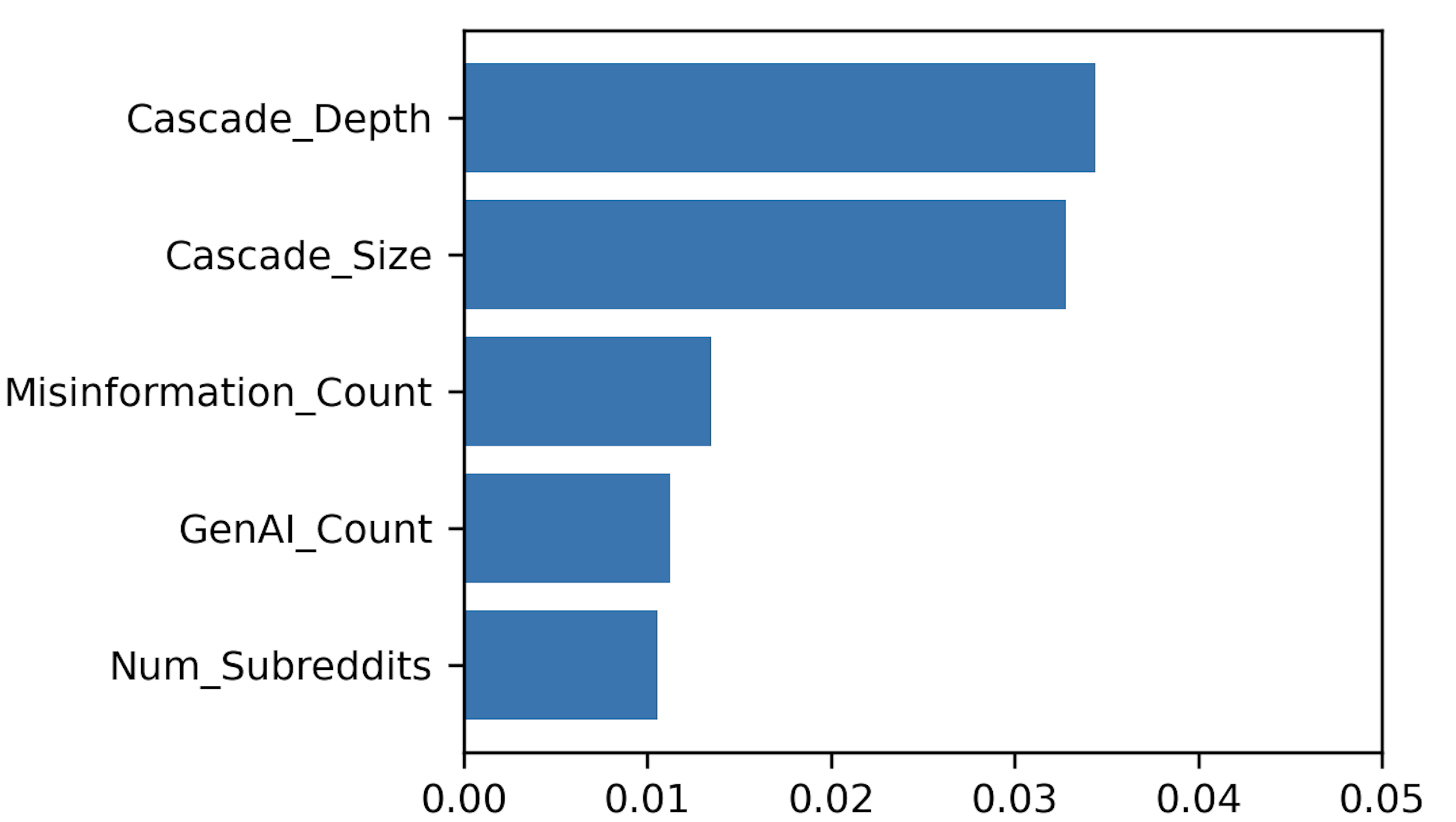}
    \caption{}
    \label{fig:rq2}
  \end{subfigure}
  \caption{Top 5 Mean SHAP Contribution for Virality Prediction at the (a) Post-level (b) Cascade-level (image, text) (c) Cascade-level (image, text, cascade-dynamics). GenAI and misinformation counts per cascade are by far the strongest virality predictors, outpacing any other image or text feature. Cascades rich in pseudo-labeled content spread faster, persist longer, and cross community boundaries, posing a moderation challenge within Reddit communities.}
\end{figure*}

\subsubsection{Virality Prediction}
We assess how well content-based features predict post (VAI) and cascade-level virality (structural virality). We compare pure text, pure image, and combined text+image features across the four models.

\subsubsection{Content vs. Context Features}
Next, we evaluate the predictive power of content features (text and image) versus context features (cascade metrics, excluding those directly tied to virality such as depth) at solely the cascade-level. To construct cascade-level text and image features, we average all post-level text and image features within a cascade.

\subsection{Results}

\subsubsection{RQ1: How do visual and textual features affect virality at the individual post and cascade levels?}

At the post level, content features yielded strong predictive performance (accuracy of 87 \%, macro-F1 of 0.78, AUC of 0.83), indicating that our models can distinguish viral posts using only image and text cues. Visual features dominated: feature-importance analysis (Fig.~\ref{fig:rq1_post}) revealed image noise as the single most influential predictor. This suggests that minor visual imperfections or artifacts may boost engagement by evoking a sense of authenticity or humor. Prior work supports this, noting that “authentic” or non-polished visuals often resonate more with viewers~\cite{khosla2014makes, dezaImageVirality2015}. Other strong visual predictors included Laplacian variance (image sharpness), with sharper images correlating with higher virality—likely due to enhanced cognitive fluency~\cite{isola2011makes}—and Error Level Analysis (ELA) scores, indicating detectable digital manipulation, which also positively associated with share rates~\cite{farid2009image}. On the textual side, posts embedding positive sentiment or explicit calls-to-action (e.g., “share this!”) performed significantly better, echoing prior findings on emotional engagement and direct prompts as sharing drivers~\cite{berger2012makes, zarrella2009science}.

At the cascade level, content features proved even more predictive. Our cascade model achieved accuracy of 89\% and an AUC of 0.998, showing that posts which lead to extensive cascades—i.e.\ sharing threads that grow into hundreds of reshares spanning multiple subreddits and time intervals—have highly distinctive content signatures. The top predictor was misinformation prevalence: posts containing demonstrably false or misleading information were far more likely to spark these large-scale diffusion events (Fig.~\ref{fig:rq1_cascade}), reinforcing findings by Vosoughi \emph{et al.}~\cite{vosoughi2018spread}. AI-generated imagery was the next strongest feature, suggesting that deepfake-style visuals lead to curiosity or alarm, driving broad resharing across the community~\cite{drolsbachMisinformation2025}. Notably, some post-level cues—such as positive sentiment—remained significant, implying that emotional tone amplifies both immediate and extended sharing.

Thus, RQ1 emphasizes how post-level virality favors attention-grabbing features like image sharpness, authentic imperfections (noise), and direct textual cues that prompt quick engagement. In contrast, cascade-level virality is driven by content that sustains public interest over time. Effective mitigation will therefore require both rapid flagging of attention-grabbing posts and sustained monitoring of misleading or synthetic content to prevent large-scale cascades.

\begin{table*}[ht]
\centering
\caption{Post‐level statistics by Misinformation and GenAI classification.}
\resizebox{\textwidth}{!}{%
\begin{tabular}{|c|c|r|r|r|r|r|r|r|r|}
\hline
\textbf{Misinformation} & \textbf{GenAI} & \textbf{Mean Age Days} & \textbf{Std Age Hours} & \textbf{Mean Score} & \textbf{Std Score} & \textbf{Mean Total Comments} & \textbf{Std Total Comments} & \textbf{Mean VAI} & \textbf{Std VAI} \\
\hline
False & False & 1,326 & 797  & 6,321  & 11,567 & 423   & 1,116 & 0.55 & 1.42 \\
False & True  & 1,313 & 778  & \underline{1,809}  & 2,634  & \underline{204} & 822   & \underline{0.17} & 0.37 \\
True  & False & 1,390 & 676  & \textbf{35,567} & 38,370 & \textbf{2,433} & 4,453 & \textbf{1.61} & 4.83 \\
True  & True  & 1,332 & 658  & \textbf{22,991} & 40,625 & \textbf{1,282} & 2,371 & \textbf{0.71} & 1.27 \\
\hline
\end{tabular}
}
\label{tab:post_stats}
\end{table*}

\begin{table*}[ht]
\centering
\caption{Cascade‐level statistics by Misinformation and GenAI classification.}
\resizebox{\textwidth}{!}{%
\begin{tabular}{|c|c|r|r|r|r|r|r|r|r|r|r|r|}
\hline
\textbf{Misinformation} & \textbf{GenAI} & \textbf{Mean Branch} & \textbf{Max Branch} & \textbf{Cascade Size} & \textbf{Cascade Depth} & \textbf{Structural Virality} & \textbf{Time to First Repost (hr)} & \textbf{Peak Repost Speed (hr)} & \textbf{Lifespan (hr)} & \textbf{\# Subreddits} \\
\hline
False & False & 0.80 & 0.86 & 2.32  & 1.32  & \underline{0.19} & \underline{126}  & \underline{3533} & \underline{1182}  & 1.001 \\
False & True  & 0.81 & 0.96 & 5.13  & 4.13  & 0.74            & 352             & 944            & 2023             & 1.000 \\
True  & False & 0.36 & 0.48 & 2.58  & 1.58  & 0.56            & \textbf{2370}   & 2473           & 2198             & 1.009 \\
True  & True  & 0.89 & 1.07 & \textbf{26.96} & \textbf{25.96} & \textbf{7.14}   & 1388            & \textbf{174}   & \textbf{13848}   & \textbf{1.106} \\
\hline
\end{tabular}
}
\label{tab:cascade_stats}
\end{table*}

\subsubsection{RQ2: When predicting cascade-level virality, how do content-based predictors compare against diffusion context-based features?}

Diffusion-based features (AUC = 0.995) outperform content-based (image/text) features alone (AUC = 0.957) in predicting cascade-level virality. While models using only content cues such as visual aesthetics, textual sentiment, and misinformation or GenAI indicators already achieve strong predictive performance (AUC = 0.957)—substantially higher than prior benchmarks, e.g., Sah et al.~(2025) who reported approximately 77\% AUC for meme virality~\cite{sah2025decoding}—incorporating diffusion signals yields significant improvements. Diffusion indicators, including rapid early reshares, cascade breadth within the initial hours, and involvement of influential users, capture immediate social dynamics crucial for predicting eventual cascade growth.

Combining both content and diffusion features provides the best predictive accuracy (AUC = 0.998), highlighting their complementary strengths. For example, visually compelling or emotionally engaging posts that also show strong initial resharing patterns have the highest likelihood of extensive virality. Furthermore, content markers such as GenAI-generated visuals and misinformation consistently rank among the strongest predictors (Fig.~\ref{fig:rq2}), underscoring moderation challenges as these types of content rapidly spread and cross community boundaries. This underscores the necessity of integrating content quality assessments with real-time diffusion dynamics to effectively predict and manage large-scale virality events.

\subsubsection{RQ3: What post‐specific features characterize pure misinformation, pure AI‐generated, and mixed‐flag classifications at the post and cascade level?}

At the post level (Table~\ref{tab:post_stats}), pure misinformation (misinfo = True, GenAI = False) drew the highest engagement with a mean score of 35,567, comments total of 2,433, and VAI 1.61. Mixed-flag posts (True/True) received moderate attention (score 22,991; comments 1,282; VAI 0.71), while pure AI-generated content (False/True) had the lowest (score 1,809; comments 204; VAI 0.17). Unlike Sah et al.\ (2025), who found similar engagement for AI and misinformation, our results suggest text-based misinformation drives attention most, with AI imagery offering only modest amplification.

At the cascade level (Table~\ref{tab:cascade_stats}), mixed-flag cascades (True/True) far outperformed others in spread and longevity. They had the highest mean cascade size of 26.96, depth of 25.96, and structural virality of 7.14, compared to pure misinformation (mean cascade size: 2.58, depth: 1.58, structural virality: 0.56) and pure AI (mean cascade size: 5.13, depth: 4.13, structural virality: 0.74). Mixed-flag content also reached its peak resharing speed fastest (174 hours) and had the longest lifespan (13,848 hours), while pure misinformation cascades spread the slowest (e.g., 2,370 hours to first repost). This extends findings from Wang et al.\ \cite{wang2023imagesCOVID}, who noted that misinformation lasts longer, by showing that AI-enhanced misinformation spreads faster and further.

RQ3 reveals a two-stage virality dynamic: text-based misinformation attracts initial engagement, AI-generated imagery accelerates early resharing, and their combination yields exceptionally viral cascades. This underscores the importance of modeling multimodal flag interactions when forecasting both immediate popularity and long-term spread.

\section{Conclusion}
This study examines how images (particularly AI-generated images) drive virality and misinformation on Reddit. We analyze a dataset of 5,660 posts across diverse subreddits to uncover how visual features contribute to the spread of AI-generated and manipulated content. These findings complement prior text-focused work, offering a more holistic understanding of online virality.

Future research could expand this analysis to other social media platforms or investigate specific image characteristics such as visual tropes or advanced deepfake techniques. Another important extension involves studying user responses to visual misinformation, including comment sentiment, perceived trustworthiness, and variation across demographic groups. Ultimately, a nuanced understanding of how images drive online engagement is essential for building defenses against the evolving landscape of manipulated media.

\section*{Acknowledgment}
The authors gratefully acknowledge Professor Adam Wierman for his invaluable support, insightful guidance, and continued encouragement throughout this research.

\bibliographystyle{IEEEtran}
\bibliography{biblio}

\end{document}